\documentclass[12pt]{article}
\usepackage{amsmath,amsthm,amsfonts,amssymb}
\def\lan{\langle}
\def\ran{\rangle}

\def\Diff{{\mathrm {Diff}}}

\def\Vir{{\mathrm {Vir}}}

\def\a{\alpha}

\def\epsilon{\varepsilon}

\def\phi{\varphi}

\def\A{{\cal A}}
\def\B{{\cal B}}
\def\O{{\cal O}}

\def\Z{{\mathbb Z}}

\def\R{{\mathbb R}}

\title{{\bf Conformal Field Theory \\ and Operator Algebras}}
\author{
{\sc Yasuyuki Kawahigashi}\footnote{Supported in part by JSPS.}\\
Department of Mathematical Sciences\\
University of Tokyo, Komaba, Tokyo, 153-8914, Japan\\
e-mail: {\tt yasuyuki@ms.u-tokyo.ac.jp}}

\begin{document}
\date{}
\maketitle

\begin{abstract}
We review recent progress in operator algebraic approach to
conformal quantum field theory.  Our emphasis is on
use of representation
theory in classification theory.  This is based on a series of 
joint works with R. Longo.
\end{abstract}

\section{Introduction}

A mathematically rigorous approach to quantum field theory based
on operator algebras is
called an {\sl algebraic quantum field theory}.  It has a long
history since pioneering works of Araki, Haag, Kastker.  (See
\cite{H} for a general treatment of algebraic quantum field theory.)
This theory works on Minkowski spaces on any spacetime
dimension, and there have been some recent
results on curved spacetimes or
even noncommutative spacetimes.  In the case of $1+1$-dimensional
Minkowski space with higher spacetime symmetry, {\sl conformal
symmetry}, we have {\sl conformal field theory} and there we have
seen many new developments in the recent years,
so we survey such results here. 
Our emphasis is on representation
theoretic aspects of the theory and we make various comparison
with another mathematically rigorous and more recent
approach to conformal field
theory, that is, theory of vertex operator algebras.

Roughly speaking, a mathematical study of quantum field theory is
a study of Wightman fields, which are certain type of
operator-valued distributions on a spacetime
with covariance with respect to a given spacetime symmetry group.
We have mathematically rigorous axioms for such Wightman fields, 
but they involve distributions and unbounded operators, so these
cause various kinds of technical difficulty.  In contrast,
in the algebraic quantum
field theory, our fundamental object is a {\sl net of von Neumann
algebras} of bounded linear operators on a Hilbert space.
(See \cite{Ta} for general theory of von Neumann algebras.)
Technical problems on definition domains of
unbounded operators do not arise in this approach.

A basic idea is as follows.  Suppose we have a Wightman field
$\Phi$ on a spacetime.  Fix a bounded region $\O$ in the
space time and consider a test function $\phi$ with support
contained in $\O$.  Then the pairing
$\lan \Phi,\phi\ran$ produces an (unbounded) operator.  We
have many $\Phi$ and $\phi$ for a fixed $\O$ and obtain many
unbounded operators from such pairing.  Then we consider a von 
Neumann algebra of {\sl bounded} linear operators on this Hilbert
space generated by these unbounded operators.   (For example,
if we have a self-adjoint unbounded operators, we consider its
spectral projections which are obviously all bounded.  In this
way, we deal with only bounded operators.)  
This is regarded as a von Neumann algebra generated by
{\sl observables} in the spacetime region $\O$.  A
von Neumann algebra is an algebra of bounded linear operators
which is closed under the adjoint operation and the strong
operator topology.  In this way, we have a family $\{\A(\O)\}$
of von Neumann algebras on the same Hilbert space parameterized
by spacetime regions.  Since the spacetime regions make a net
with respect to the inclusion order, we call such a family 
a net of von Neumann algebras.  Now we forget Wightman fields
and consider only a net of von Neumann algebras.  We have
some expected properties for such nets of von Neumann algebras
from a physical consideration, and now we use these properties
as {\sl axioms}.  So our mathematical object is a net of
von Neumann algebras subject to certain set of axioms.  Our
mathematical aim is to study such nets of von Neumann algebras.

\section{Conformal Quantum Field Theory}

We first explain formulation of full conformal quantum
field theory on the  $1+1$-dimensional Minkowski space
in algebraic quantum field theory.  
As a spacetime region $\O$ above, it is enough
to consider only open rectangles $\O$ with edges parallel
to $t=\pm x $ in $(1+1)$-dim Minkowski space.
In this way, we get a
family $\{\A(\O)\}$ of operator algebras
parameterized by spacetime regions $\O$ (rectangles).
In order to realize conformal symmetry, we have to make
a partial compactification of the $1+1$-dimensional Minkowski space.
If two rectangles are spacelike separated, then we have no 
interactions between them even at the speed of light, so our
axiom requires that the corresponding two von Neumann algebras
commute with each other.  This is the {\sl locality
axiom}.  Since this is not our main object in this paper,
we omit details of the other axioms.  See \cite{KL2} for full details.  

Next we briefly explain that boundary conformal field theory can
be handled within the same framework.  Now we consider the
half-space $\{(x,t)\mid x>0\}$ in the $1+1$-dimensional Minkowski
space and only rectangles $\O$ contained in this half-space.
In this way, we have a similar net of von Neumann algebras
$\{\A(\O)\}$ parameterized with rectangles in the half-space.
See \cite{LR2} for full details of the axioms.

If we have a net of von Neumann algebras over the $1+1$-dimensional
Minkowski space, we can {\sl restrict} the net of von Neumann algebras
to two {\sl chiral} conformal field theories on the light cones
$\{x=\pm t\}$.  In this way, we have two nets of von Neumann
algebras on the compactified $S^1$ as description of two {\sl
chiral} conformal field theories.  Since this net is our main 
mathematical object in this article, we give a full set of
axioms.  (See \cite{KL2} for details of this ``restriction''
procedure.)

Now our ``spacetime'' is $S^1$ and a ``spacetime region''
is an interval $I$, which means a non-empty, non-dense open
connected subset of $S^1$.
We have a family $\{\A(I)\}$ of von Neumann algebras on a fixed
Hilbert space $H$.  These von Neumann algebras are simple and
such von Neumann algebras  are called {\sl factors}, so
the family $\{\A(I)\}$ satisfying the axioms below is
called a {\sl net of factors} (or an {\sl irreducible
local conformal net
of factors}, strictly speaking).
Actually, the set of intervals on $S^1$ is not
directed with respect to inclusions, so the terminology 
{\sl net} is not mathematically appropriate, but is widely
used.  

\begin{enumerate}
\item (isotony)
For intervals $I_1\subset I_2$, we have $\A(I_1)\subset \A(I_2)$.
\item (locality) For intervals $I_1, I_2$ with
$I_1\cap I_2=\varnothing$, we have
$[\A(I_1), \A(I_2)]=0$
\item (M\"{o}bius covariance)
There exists a strongly 
continuous unitary representation $U$ of $PSL(2, \R)$ on $H$
satisfying $U(g)\A(I) U(g)^*= \A(gI)$ for any $g\in PSL(2,\R)$
and any interval $I$.
\item (positivity of energy)
The generator of the one-parameter rotation subgroup of $U$,
called the {\sl conformal Hamiltonian}, is positive. 
\item (existence of the vacuum) There exists a unit 
$U$-invariant vector $\Omega$ in $H$,
called the {\sl vacuum vector}, and the von Neumann algebra
$\bigvee_{I\in S^1}\A(I)$ generated by all $\A(I)$'s
is $B(H)$.
\item (conformal covariance) There exists a projective unitary 
representation $U$ of $\Diff(S^1)$ on $H$ extending the unitary 
representation of $PSL(2,\R)$ such that for all intervals $I$,
we have
\begin{eqnarray*}
U(g)\A(I) U(g)^* &=& \A(gI),\quad  g\in\Diff(S^1), \\
U(g)AU(g)^* &=&A,\quad A\in\A(I),\ g\in\Diff(I'),
\end{eqnarray*}
where $\Diff(S^1)$ is the group of orientation-preserving 
diffeomorphisms of $S^1$ and $\Diff(I')$ is the group of 
diffeomorphisms $g$ of $S^1$ with $g(t)=t$ for all $t\in I$.
\end{enumerate}

The isotony axiom is natural because we have more test functions
(or more observables) for a larger interval.  The locality axiom
takes this simple form on $S^1$.
The choice of the spacetime symmetry is not unique, and we can
use the Poincar\'e symmetry on the Minkowski space or 
the M\"obius covariance on $S^1$, for example, but in the
{\sl conformal}
field theory, we use {\sl conformal} symmetry, which means
diffeomorphism covariance as above.
This set of axioms imply various nice conditions such as
the Reeh-Schlieder property, the Bisognano-Wichmann property
and the Haag duality.  See \cite{KL1} and references there for
details.

In the usual situation, all the von Neumann algebras $\A(I)$ are
isomorphic to the so-called Araki-Woods type III$_1$ factor
for all nets $\A$ and all intervals $I$.  So each von Neumann
algebra does not contain any information about the conformal
field theory, but it is the relative position of the von Neumann
algebras in the family that encodes the physical information of
the theory.  (It is similar to subfactor theory of Jones where
we study a relative position of one factor in another.)

At the end of this section, we compare our formulation of conformal
quantum field theory with another mathematically rigorous approach,
theory of vertex operator algebras.
A vertex operator algebra is an algebraic axiomatization of
Wightman fields on $S^1$, called vertex operators.
If we have an operator valued
distribution on $S^1$, its Fourier expansion should give countably
many (possibly unbounded) operators as the Fourier coefficients.
Under the so-called state-field correspondence, any vector in
the space of ``states'' should give an operator-valued distribution,
a quantum ``field'', and its Fourier expansion gives countably
many operators.  In this way, one vector should give countably
many operators on the space of these vectors.  In other words,
for two vectors $v,w$ we have countably many binary operations
$v_{(n)}w$, $n\in\Z$, the action of the $n$-th operator given
by $v$ on $w$.  An axiomatization of this idea gives a
notion of {\sl vertex operator algebra}.  (See \cite{FLM}
for a precise definition.  There is a slightly weaker notion of
a {\sl vertex algebra}.  See \cite{K} for its precise definition
and related results.)  In theory of vertex operator algebra,
one considers a vector space of states without an inner product
and even when we have a positive definite inner product, one 
considers this vector space without completion.  Here in 
comparison to nets of factors, we are interested in the case
where we have positive definite inner products on the spaces
of states.  We say that such a vertex operator algebra is {\sl unitary}.

Both of one (unitary)
vertex operator algebra and one net of factors should describe
one chiral conformal field theory.  So unitary vertex operator
algebras and
nets of factors should be in a bijective correspondence, at least
under some ``nice'' additional
conditions, but no general theorems have been known
for such a correspondence, though there is a recent progress due to
S. Carpi and M. Weiner.
However, if we have one construction or an idea on one side,
we can often
``translate'' it to the other side, though it can be highly
non-trivial from a technical viewpoint.  Fundamental sources
of constructions for vertex operator algebras are
affine Kac-Moody algebras and integral lattices.
The corresponding constructions for nets of factors have
been done by A. Wassermann \cite{W} and his students, and
Dong-Xu \cite{DX}, respectively, after the initial construction
of Buchholz-Mack-Todorov \cite{BMT}.  If we have examples
with some nice properties, we canoften  construct new
examples from them, and as such methods of constructions of
vertex operator algebras, we have
simple current extensions, the coset 
construction, and the orbifold construction.  The
simple current extensions for nets of factors are simply
crossed products by DHR-automorphisms and easy to realize.
(See the next section for a notion of DHR-endomorphisms.)
The coset and orbifold constructions for nets of factors
have been studied in detail by F. Xu \cite{X3,X4,X5}.

For nets of factors, we have introduced a new construction
of examples in \cite{KL1} based
on Longo's notion of $Q$-systems \cite{L2}.  Further
examples have been constructed by Xu \cite{X8} with
this method.  This can be
translated to the setting of vertex operator algebras,
as we will see in this article later.

\section{Representation Theory}

An important tool to study nets of factors is a representation theory.
For a net of factors $\{\A(I)\}$,
all the algebras $\A(I)$ act on the initial Hilbert space $H$
from the beginning, but we also consider their representations
on another Hilbert space, that is, a family $\{\pi_I\}$ of
representations $\pi_I: {\A}(I)\to B(K)$, where $K$
is another Hilbert space, common for all $I$.  For
$I_1\subset I_2$, we must have that the restriction of $\pi_{I_2}$
on $\A(I_1)$ is equal to $\pi_{I_1}$.  
The representation on the initial Hilbert space is called
the {\sl vacuum representation} and plays a role of a trivial 
representation.  We also have to take care of
the spacetime symmetry group when we consider a representation,
but this part is often automatic
(see \cite{GL1}), so we now ignore it for simplicity.
See \cite{GL1} for a more detailed treatment.  Note that
a representation of a net of factors
is a counterpart of a module over a vertex operator algebra.

Notions of irreducibility and a direct sum for
such representations are easy to formulate.
Non-trivial notions are dimensions and tensor products.  
Each representation $\{\pi_I\}$ is in a bijective correspondence
to a certain {\sl endomorphism} $\lambda$ of an infinite dimensional
operator algebra, called a Doplicher-Haag-Roberts (DHR)
endomorphism \cite{DHR,FRS}, and we can restrict $\lambda$ to a single
factor $\A(I)$ for an arbitrarily but fixed interval $I$.
Then $\lambda(\A(I))\subset \A(I)$ is a subfactor
and we have its Jones index \cite{J}.  (See \cite{EK,O1,P} for general
theory of subfactors.)  The square root  of this Jones index
plays the role of the {\sl dimension}
of the representation \cite{L1}.  In algebraic quantum field
theory, such a dimension was called a {\sl statistical dimension},
and it is analogous to a quantum dimension in the theory of
quantum groups.  It is a positive real numbers in the interval
$[1,\infty]$.  We can also compose endomorphisms
and this composition gives the correct notion of {\sl tensor products}.
We then get a {\sl braided} tensor category as in \cite{FRS}.

In representation theory of a vertex operator algebra (and also
a quantum group), it sometimes happens that we have only finitely
many irreducible representations.
Such finiteness is often called {\sl rationality}, possibly with
some extra assumptions on some finite dimensionality.
This also plays an important role in theory of quantum
invariants in low dimensional topology.
In \cite{KLM}, we have introduced an operator algebraic condition
for such rationality for nets of factors as follows and we
called it {\sl complete rationality}.  
We split the circle into four intervals $I_1, I_2, I_3, I_4$ in this
order, say, counterclockwise.
Then complete rationality is given by the finiteness of the
Jones index for a subfactor
$\A(I_1)\vee\A(I_3)
\subset (\A(I_2)\vee\A(I_4))'$
where $'$ means the commutant, together with the {\sl split
property}.  The split property is known to hold if the {\sl vacuum
character}, $\sum_{n=0}(\dim H_n) q^n$, is convergent for $|q|<1$
by \cite{DLR}, so it usually holds and is easy to verify.
(Here $H=\bigoplus_{n=0}^\infty H_n$ is the eigenspace
decomposition of the original Hilbert space for the positive
generator of the rotation group.  So this convergence property
can be verified simply by looking at the Hilbert space, not 
the von Neumann algebras.)  In the original definition of
complete rationality in \cite{KLM},
we required another condition called strong additivity, but
it was proved to be redundant by Longo-Xu \cite{LX}.
We have proved in \cite{KLM} that this
complete rationality implies that we have 
a {\sl modular} tensor category as a representation category
of $\{\A(I)\}$.  A modular tensor category produces a $3$-dimensional
topological quantum field theory.  (See \cite{T} for general theory
of topological quantum field theory.)
The $SU(N)_k$-net of Wassermann
has been shown to be completely rational by \cite{X2}.

We now introduce an important notion of {\sl $\alpha$-induction}.
For an inclusion of nets of factors,
$\A(I)\subset \B(I)$, we have
an induction procedure analogous to the group representation.
So from a representation of the smaller net $\A$, we would like
to construct a representation of the larger net $\B$, but what
we actually obtain is not a genuine representation of the
larger net $\B$ in general, and is something weaker called
{\sl solitonic}.  This induction procedure is called the
$\alpha$-induction and depends a choice of braiding, so
we write $\alpha^+$ and $\alpha^-$.
This was first defined in Longo-Rehren
\cite{LR1} and studied in detail in Xu \cite{X1}.
Then B\"ockenhauer-Evans \cite{BE} made a further study, and
\cite{BEK1,BEK2} unified this study with Ocneanu's graphical
method \cite{O2}.
The intersection of the irreducible endomorphisms appearing
in the images of $\alpha^+$-induction and
$\alpha^-$-induction gives the true representation category
of $\{\B(I)\}$ if $\A$ is completely rational by \cite{BEK1,KLM}.

This $\alpha$-induction opens an important and new connection with
theory of {\sl modular invariants}.
A modular tensor category produces a unitary representation $\pi$
of $SL(2,{\mathbb Z})$ through its braiding as in \cite{R}, and
its dimension
is the number of irreducible objects.  So a completely rational
net of factors produces such a unitary representation.
(Note that our representation of $SL(2,{\mathbb Z})$ comes from
the braiding structure, not from the action of this group
on the characters through change of variables
$\tau\mapsto \displaystyle\frac{a\tau+b}{c\tau+d}$, though
in all the ``nice'' known examples, these two representations
coincide.  See \cite{KL3} for a discussion on this matter.)

It has been proved in \cite{BEK1} that
the matrix $(Z_{\lambda,\mu})$ defined by
$$Z_{\lambda,\mu}=\dim {\mathrm{Hom}} (\alpha^+_\lambda,\alpha^-_\mu)$$
is in the commutant of the representation $\pi$, using Ocneanu's
graphical calculus \cite{O2}.   Such a matrix
$Z$ is called a {\sl modular invariant}, and we have only finitely
many such $Z$ for a given $\pi$.  For any completely rational
net $\{\A(I)\}$, any extension
$\{\B(I)\supset\A(I)\}$ produces such $Z$.  Matrices
$Z$ are certainly much easier to classify than extensions and this
is a source of classification theory in the next section.

\section{Classification Theory}

For a net of factors, we can naturally define a {\sl central charge}
and it is well-known to take discrete values $1-6/m(m+1)$, $m=3,4,5,\dots$,
below $1$ and all values in $[1,\infty)$ by \cite{FQS,GKO}.  We have the
{\sl Virasoro net} $\{{\mathrm{Vir}}_c(I)\}$ for each such $c$ and
it is the operator algebraic counterpart of
the Virasoro vertex operator algebra with the same $c$.  Any net of factors
$\{\A(I)\}$ with central charge $c$
is an extension of the Virasoro net with the same central charge
and it is automatically completely rational if $c<1$,
as shown in \cite{KL1}.  So we can apply
the above theory and we get the following complete classification
list for the case $c<1$ as in \cite{KL1}.

\begin{enumerate}
\item The Virasoro nets $\{{\mathrm{Vir}}_c(I)\}$ with $c<1$.
\item The simple current extensions of the Virasoro nets with index 2.
\item Four exceptionals at $c=21/22$, $25/26$,
$144/145$, $154/155$.
\end{enumerate}

The unitary representations of $SL(2,\Z)$ for the Virasoro nets
are the well-known ones, and all the modular invariants for
these have been classified by \cite{CIZ}.  Our result shows
that each of the so-called type I modular invariants in the
classification list of \cite{CIZ} corresponds to a net of factors
uniquely.
They are labeled with pairs of $A$-$D_{2n}$-$E_{6,8}$ Dynkin diagrams with
Coxeter numbers differing by 1.
Three in (3) of the above list have been identified with coset
models, but the remaining one does not seem to be related to any other
known constructions.  This is constructed with
``extension by Q-system''.
Xu \cite{X8} recently applied this construction to many other coset models
and obtained infinitely many new examples
based on \cite{X7}, called {\sl mirror extensions}.
Classification for the case $c=1$ has been also done under some
extra assumption \cite{C,X6}.

This classification theorem also implies a classification of certain
types of vertex operator algebras as follows.

Let $V$ be a (rational) vertex operator algebra and $W_i$ be its
irreducible modules.  We would like to
classify all vertex operator algebras arising from putting a 
vertex operator algebra structure on
$\bigoplus_i n_i W_i$ and using the same Virasoro element as $V$,
where $n_i$ is multiplicity and $W_0=V$, $n_0=1$.  From a
viewpoint of tensor category, this classification
problem of extensions of a vertex operator algebras is the ``same'' as
the classification problem of extensions of
a completely rational net of factors, as shown in \cite{HKL}.

So the above classification theorem of local conformal
nets implies a classification
theorem of extensions of the Virasoro vertex operator algebras
with $c<1$ as above, and we obtain the same classification list.
That is, besides the Virasoro vertex operator algebras themselves,
we have their simple current extensions, and four exceptionals
at $c=21/22, 25/26, 144/145, 154/155$.  With the usual notation 
of $L(c, h)$ for a module with central charge $c$ and conformal
weight $h$ of the Virasoro vertex operator algebras with $c<1$,
the four exceptionals are listed as follows.
\begin{enumerate}
\item $L(21/22,0)\oplus L(21/22,8)$.  It has
15 irreducible representations and has two coset realizations,
from $SU(9)_2\subset (E_8)_2$ and
$(E_8)_3\subset (E_8)_2\otimes (E_8)_1$.
\item $L(25/26,0)\oplus L(25/26,10)$.  It has
18 irreducible representations and has a
coset realization from $SU(2)_{11}\subset SO(5)_1\otimes SU(2)_1$.
\item $L(144/145,0)\oplus L(144/145,24)\oplus
L(144/145,78)\oplus L(144/145,189)$.  It has
28 irreducible representations and no coset realization has been known.
\item $L(154/155,0)\oplus L(154/155,26)\oplus
L(154/155,84)\oplus L(154/155,203)$.  It has
30 irreducible representations and has a 
coset realization from $SU(2)_{29}\subset (G_2)_1\otimes SU(2)_1$.
\end{enumerate}

Note that it is not obvious that the representation category of
the Virasoro net $\Vir_c$ and the representation category of
the Virasoro vertex operator algebra $L(c,0)$ are
isomorphic, but as long as the
two are braided tensor category and have the same $S$- and $T$-matrices,
the arguments in \cite{KL1} work, so we obtain the above classification
result for vertex operator algebras.

Using the above results and more techniques, we can also completely
classify full conformal field theories within the
framework algebraic quantum field theory for the case $c<1$.
Full conformal field theories are given as certain nets of
factors on $1+1$-dimensional Minkowski space.  Under natural
symmetry and maximality conditions, those with $c<1$ are
completely labeled with the pairs of $A$-$D$-$E$ Dynkin diagrams with
the difference of their Coxeter numbers equal to 1, as shown 
in \cite{KL2}.  We now naturally have $D_{2n+1}$, $E_7$
as labels, unlike in the chiral case.
The main difficulty in this work lies in proving uniqueness
of the structure for each modular invariant in the
Cappelli-Itzykson-Zuber list \cite{CIZ}.  This is done
through 2-cohomology vanishing for certain tensor categories.
in the spirit of \cite{IK}.

Furthermore, using the above results and more techniques
we can also completely
classify boundary conformal field theories for the case $c<1$.
Boundary conformal field theories are given as certain nets of
factors on a $1+1$-dimensional Minkowski half-space.  Under a natural
maximality condition, these with $c<1$ are now
completely labeled with the pairs of
$A$-$D$-$E$ Dynkin diagrams with distinguished vertices having
the difference of their Coxeter numbers equal to 1, as shown
in \cite{KLPR} based on a general theory in \cite{LR2}.
The ``chiral fields'' in a boundary conformal field theory
should produce a net of factors on
the boundary (which is compactified to $S^1$) as in the operator
algebraic  approach.  Then a general
boundary conformal field theory restricts to this boundary to produce a
{\sl non-local} extension of this
{\sl chiral} conformal field theory on the boundary.

\section{Moonshine Conjecture}

The Moonshine conjecture, formulated by Conway-Norton \cite{CN},
is about mysterious relations between finite simple groups and
modular functions, since an observation due to McKay.

Today the classification of all finite simple groups is complete 
and the classification list contains 26 sporadic groups in addition
to several infinite series.  
The largest group among the 26 sporadic groups is called the
{\sl Monster} group and its order is about $8\times10^{53}$

One the one hand, the non-trivial irreducible representation 
of the Monster having the smallest dimension is 196883 dimensional.
On the other hand, the following function, called {\sl $j$-function},
has been classically studied in algebra.
\begin{eqnarray*}
j(\tau)&=&q^{-1}+744+196884q+\\
&&\quad 21493760q^2+864299970 q^3+\cdots
\end{eqnarray*}
For $q=\exp(2\pi i\tau)$, ${\mathrm{Im}}\; \tau >0$, we have modular
invariance property,
$j(\tau)=j\left(\displaystyle\frac{a\tau+b}{c\tau+d}\right)$ for
$\left(\begin{array}{cc}
a & b \\ c& d\end{array}\right)\in SL(2,{\mathbb{Z}})$, and this
is the only function, up to the constant term,
satisfying this property and starting with $q^{-1}$,

McKay noticed $196884=196883+1$, and similar simple relations for
other coefficients of the $j$-function and dimensions of irreducible
representations of the Monster group turned out to be true.
Then Conway-Norton \cite{CN} formulated
the Moonshine conjecture roughly
as follows, which has been now proved by Borcherds \cite{B} in 1992.

\begin{enumerate}
\item 
We have a ``natural'' infinite dimensional graded vector space
$V=\bigoplus_{n=0}^\infty V_n$ with some algebraic structure
having a Monster action preserving
the grading and each $V_n$ is finite dimensional.
\item
For any element $g$ in the Monster, the power series
$\sum_{n=0}^\infty ({\mathrm{Tr}}\;g|_{V_n})q^{n-1}$ is a
special function called a {\sl Hauptmodul} for
some discrete subgroup of $SL(2,{\mathbb{R}})$.
When $g$ is the identity element, the series is
the $j$-function minus constant term 744.
\end{enumerate}

For the part (1) of this conjecture,
Frenkel-Lepowsky-Meurman \cite{FLM}
gave a precise definition of ``some algebraic structure''
as a {\sl vertex operator algebra}
and constructed  a particular example $V$, which is now called
the {\sl Moonshine vertex operator algebras} and denoted by
$V^\natural$.

The construction roughly goes as follows.
In dimension 24, we have an exceptional lattice $\Lambda$
called the {\sl Leech lattice}.  Then there is a general
construction of a vertex operator algebra from a certain
lattice, and the one for the Leech lattice gives
something very close to our final object $V^\natural$.
Then we take a fixed point algebra under a natural action of
${\mathbb{Z}}/2{\mathbb{Z}}$ arising from the lattice symmetry,
and then make a simple current
extension of order 2.  The resulting vertex operator algebra
is the Moonshine vertex operator algebra
$V^\natural$.  (The final step is called a
twisted orbifold construction).  The series
$\sum_{n=0}^\infty (\dim V^\natural_n)q^{n-1}$ is indeed the
$j$-function minus constant term 744.

Miyamoto \cite{M} has a new realization of  $V^\natural$
as an extension of a tensor power of the Virasoro vertex operator
algebra with $c=1/2$, $L(1/2,0)^{\otimes 48}$
(based on Dong-Mason-Zhu \cite{DMZ}).  This kind of extension
of a Virasoro tensor power is called a {\sl framed vertex
operator algebra} as in \cite{DGH}.

We have given an operator algebraic counterpart of such
a construction in \cite{KL4}.

We realize a Leech lattice net of factors on $S^1$ as an extension
of ${\mathrm{Vir}_{1/2}}^{\otimes 48}$ using certain
${\mathbb Z}_4$-code.  Then we can perform the twisted orbifold
construction in the operator algebraic sense to obtain
a net of factors, the {\sl Moonshine net}
${\A}^\natural$.  Theory of $\alpha$-induction is used for
obtaining various decompositions.  We then get a Miyamoto-type
description of this construction, as an operator algebraic
counterpart of the framed vertex operator algebras. 
We then have the following properties.

\begin{enumerate}
\item $c=24$.
\item The representation theory is trivial.
\item The automorphism group is the Monster.
\item The Hauptmodul property (as above).
\end{enumerate}

Outline of the proof of these four properties is as follows.

It is immediate to get $c=24$.  We can show
complete rationality passes to
an extension (and an orbifold) in general
with control over the size of
the representation category, using the Jones index.  With this,
we obtain (2) very easily.  Such a net is called {\sl holomorphic}.
Property (3) is the most difficult part.  For the Virasoro
VOA $L(1/2,0)$, the vertex operator is indeed a well-behaved Wightman
field and smeared fields produce the Virasoro net
${\mathrm{Vir}}_{1/2}$.  Using this property and the fact
that $\bigcup_g g (L(1/2,0)^{\otimes 48})$ for all
$g\in{\mathrm{Aut}}(V^\natural)$
generate the entire Moonshine VOA $V^\natural$,
we can prove that the automorphism group
as a vertex operator algebra
and the automorphism group as a net of factors are indeed the same.
Then (4) is now a trivial corollary of the Borcherds theorem \cite{B}.

We note that the Baby Monster, the second largest among the 26 sporadic
finite simple groups, can be treated similarly with H\"ohn's construction
of the shorter Moonshine super vertex operator algebra.

Still, these examples are treated with various tricks case by case.
We expect
a bijective correspondence between vertex operator algebras and
nets of factors
on $S^1$ under some nice conditions.
On the side of vertex operator algebras, the most natural candidate
for such a ``nice''
condition is the $C_2$-finiteness condition of Zhu \cite{Z}
(with unitarity).
On the operator algebraic side, our complete rationality in \cite{KLM}
seems to be such a ``nice'' condition, but the actual relations between
the two notions are not clear at this moment.
The essential condition for complete rationality is the finiteness of
the Jones index arising from four intervals on the circle, and
this finiteness somehow has formal similarity to the finiteness
appearing in the definition of the $C_2$-finiteness.

At the end, we list some open problems.
The operator algebraic approach has an advantage in control of
representation theory, but is behind of theory of vertex
operator algebras in the theory of characters.

For a net of factors, we can naturally  define a notion of a
character for each representation.  But  even
convergence of these characters have not been proved in general, and
the modular invariance property, the counterpart of
Zhu's result \cite{Z}, is unknown, though we certainly expect it
to be true.  We also expect the Verlinde identity holds,
which has been proved in the context of vertex operator
algebras recently by Huang \cite{Hu}.  We would need an $S$-matrix
version of the spin-statistics theorem \cite{GL2} for nets
of factors.

\end{document}